# Recoil Analysis for Heavy Ion Beams

Fatih EKİNCİ[1], Erkan BOSTANCI[2], Özlem DAĞLI[3] and Mehmet Serdar GÜZEL[2]

fatih.ekinci2@gazi.edu.tr, ebostanci@ankara.edu.tr, ozlemdagli@gazi.edu.tr, mguzel@ankara.edu.tr
1 Physics Department, Gazi University
2 Computer Engineering Department, Ankara University,
3 Gamma Knife Unit, Gazi University

## Abstract

Given that there are 94 clinics and more than 200,000 patients treated worldwide, proton and carbon are the most used heavily charged particles in heavy ion therapy. However, there is a recent increasing trend in using new ion beams. Each heavy ion has a different effect on the target. As each heavy ion moves through the tissue, they lose enormous energy in collisions, so their range is not long. Ionization accounts for the majority of this loss in energy. During this interaction of the heavily charged particles with the target, the particles do not only ionize, but also lose energy with the recoil. Recoil occurs by atom-to-atom collisions. With these collisions, crystalline atoms react with different combinations and form cascades in accordance with their energies. Thus, secondary particles create ionization and recoil. In this study, recoil values of boron, carbon, nitrogen and oxygen beams in the water phantom were computed in the energy range of 2.0-2.5 GeV using Monte Carlo simulation and the results were compared with carbon. It was observed that there is a regular increase in the recoil peak amplitude for carbon and boron ions, unlike oxygen and nitrogen where such a regularity could not be seen. Moreover, the gaps in the crystal structure increased as the energy increases.

## 1. Introduction

Hadron beams have been used in radiation oncology for a long time due to their superior physical and biological properties compared to conventional high-energy photon beams [1, 2]. Therefore, the number of particle treatment facilities has increased significantly in the last few years, despite their cost and technological difficulties faced [3]. Protons are currently used in more than 60 facilities around the world, there are 16 centers in clinical treatments in Europe and many are under construction [4]. Based on the excellent clinical results achieved with carbon ion beams in Japan, four carbon ion therapy centers have been established in Europe in the last decade [4]. More recently, researchers have also focused on newer types of particles other than protons and carbon ions, namely helium and oxygen [5-9]. The clinical outcome of particle therapy depends on dosimetric accuracy, including accurate dose calculations and beam delivery, respectively, as well as various clinical aspects. Much of the clinical experience in particle beam therapy to date has been achieved through radiotherapy treatment planning and dose calculations based on semi-analytical pencil beam algorithms [10].

With regard to dosimetric accuracy, general purpose Monte Carlo (MC) simulations are considered the "gold standard" [11]. MC method is a statistical simulation technique developed for solving mathematical problems where finding an analytical solution is difficult. Simulation systems developed on this technique follow the traces of each particle traveling through matter one by one, based on the assumption that the quantities describing particle interactions have certain probability distributions. For many particles, quantities such as flux, energy loss and absorbed Linear Energy Transfer (LET) are recorded and average values are computed for these distributions [12]. TRansport of Ions in Matter (TRIM) simulation software developed using MC technique allows computation of all ion interactions within the target. Input parameters such as the ion type, energy, number of ions and related probabilities as well as target phantom properties including shape and material can be provided to the software. The software records all types of computed fields and can display them as needed. TRIM can compute all kinetic events and 3D distribution of ions related to energy loss processes of ions such as target damage, scattering, ionization, phonon generation and recoil. All target atom cascades in the target can be tracked in detail. The software also allows step by step analysis of all tracks [13].

The authors have identified a gap in the calculation of the recoil profiles of the heavy ions used in heavy ion therapy and the comparison of their values in the current literature. The recoil values were compared in order to find out which ions can be more effective in the treatment of tumors that are close to critical tissues. The main purpose of this study is to reveal recoil processes when all interaction processes are considered since these processes can significantly affect the efficiency in the whole heavy ion process except ionization. For this reason, the recoil values of heavy ions already used and planned to be used were evaluated using the TRIM simulation software.

The rest of the study is structured as follows: Section 2 describes the methods used in the study, followed by Section 3 where findings are presented. Section 4 presents a detailed discussion on the findings and the paper is concluded in Section 5.

## 2. Methods

TRIM can be used to compute detailed results of collision tables and collision cascades of each ion with atoms of the target. First, the current ion energy and depth is given and then the energy loss of the ion with the target electrons, *i.e.* the electronic stopping power called "SP", is given by the unit eV/Angstrom. Each cascade causes displacement collisions, gap generation, secondary collisions, and intermediate atom production. The number of displacement collisions indicates how many target atoms are in motion at energy above the energy of displacement. Another feature in the table is Target Gaps dedicated to the gap left behind when the rebound atom leaves its original location. A moving atom strikes a fixed target atom and transfers more than its displacement energy. If it does not have enough energy to move the first atom forward after the collision and it is the same element as the atom it hits, then no vacuum is formed [13,14].

Gaps occur in the crystal structure when the bullet atoms collide with the target atoms. The damping place of a moving rebound atom may be somewhat far from the gap it leaves [13]. When hadrons interact with the target material, only ionization does not occur. Just as bullet atoms interact with material electrons, atomic-sized collisions can also occur. Target atoms are displaced by this interaction and gaps occur in the target. In order to understand such interactions, it is necessary to explain Displacement Energy and Molecular Bonding Energy [12, 14].

Displacement energy is the energy required by a rebound atom to overcome the target's molecule binding force and remove multiple atomic gaps from their original positions. A bullet assumes that the atom has atomic number $Z_1$ and energy E and has the probability of a collision with an atom with atomic number $Z_2$ in the target. Let the energy of the bullet ion be $E_1$ and the energy of the atom hit $E_2$ after the collision. $E_d$, energy of displacement; $E_b$ becomes the binding energy of a molecular atom and $E_f$ becomes the final energy of a moving atom small enough to be considered to be at rest [13, 15].

There are 3 possibilities in the recoil reaction:
1. First, a displacement occurs at $E_2 > E_d$ (enough energy is supplied to the target atom to detach from its position). If $E_1 > E_d$ and $E_2 > E_d$ (both atoms have enough energy to leave their position) a vacuum is formed. Both atoms become mobile atoms. The $E_2$ energy of the $Z_2$ atom is reduced by $E_b$ before another collision. If $E_2 < E_d$ then the atom being hit does not have enough energy. Atom returns to its original position by emitting E2 energy as a photon [13].
2. Second, if $E_1 < E_d$ and $E_2 > E_d$ and $Z_1 = Z_2$, then the incoming atom is captured. $E_1$ oscillates as a phonon, and this collision is called a displacement collision. This type of impact is common on single element targets with large recoil cascades. If $E_1 < E_d$ and $E_2 > E_d$ and $Z_1 \neq Z_2$, $Z_1$ becomes an interstitial atom that forms or occupies gaps [13].
3. Finally, if $E_1 < E_d$ and $E_2 < E_d$, $Z_1$ becomes a transition atom and $E_1 + E_2$ energy is released as a photon. If there are several different elements in a target and each has a different energy of displacement, then $E_d$ will change for each atom of the cascade hitting different target atoms [13].

As with photon radiotherapy, the most important problem for hadron therapy is whether the desired dose can be administered to the patient. For this, an attempt is made to determine and calibrate the correct dose using the water phantom before the patient is treated [13]. Water is the most important medium used in medical physics due to its similarity to human tissues in terms of atomic weight and density. Reliability of stopping power calculations for water and accurate calculation of dose distribution mean accurate treatment doses for patients since the main component of the human body is considered water. In hadron therapy applications, similar to photon radiotherapy, dose distribution is controlled by tissue equivalent phantoms (such as water phantoms).

In this respect, the shape and design of the phantom structure to be used are important parameters for a simulation environment. There are various types of phantoms used for different body planning in literature [16]. In this study, a cylindrical water phantom is used (see Figure 1). In the TRIM simulation software, the atomic density of the water is 10,0222 atoms/cm³ and its density is 1g/cm³. The atomic combination ratios of hydrogen (H) and oxygen (O) atoms forming the water molecule are given as 66.6% and 33.3%, respectively. Similarly, the mass association ratios are given as 11.1% for H and 88.8% for O. The TRIM simulation program determined these percentages according to the ICRU-276 report [17].

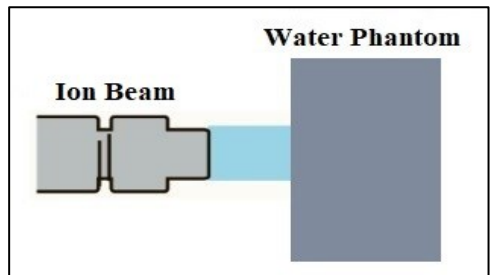

**Figure 1.** Water phantom

In this study, recoil values of carbon, boron, nitrogen and oxygen beams were obtained by increasing the energy in steps of 0.1 GeV in the energy range of 2.0-2.5 GeV. Results were compared using the Bragg curve of the back beam of the carbon beam and the recoil. In the calculations, the beams of carbon and boron were sent to the target in such a way that statistical deviations were in acceptable ranges.

## 3. Findings

In order to test the accuracy of the calculations in this study, the Bragg curves of carbons with 1.6, 2.4 and 3.0 GeV energies in the water phantom were compared with the studies in the literature [18-22]. Based on the results obtained, it was observed that the difference was approximately 4.5% (within acceptable limits i.e. ≤5% in the medical field). The inhomogeneity effects and MC-based probabilities may sometimes result in such discrepancies, though they were within acceptable limits.

While heavy ions lose 99.8% of their energy through ionization in the target, they lose 0.02% with recoils. Recoil energies are depicted for the C, B, N and O ions for 2 GeV in Figures 2 and 3. Recoil interaction is the most important factor that changes the direction of the bullet particle, causing deviations in the direction of the advance through the target. These deviations are of great concern in tumor treatment near critical tissues.

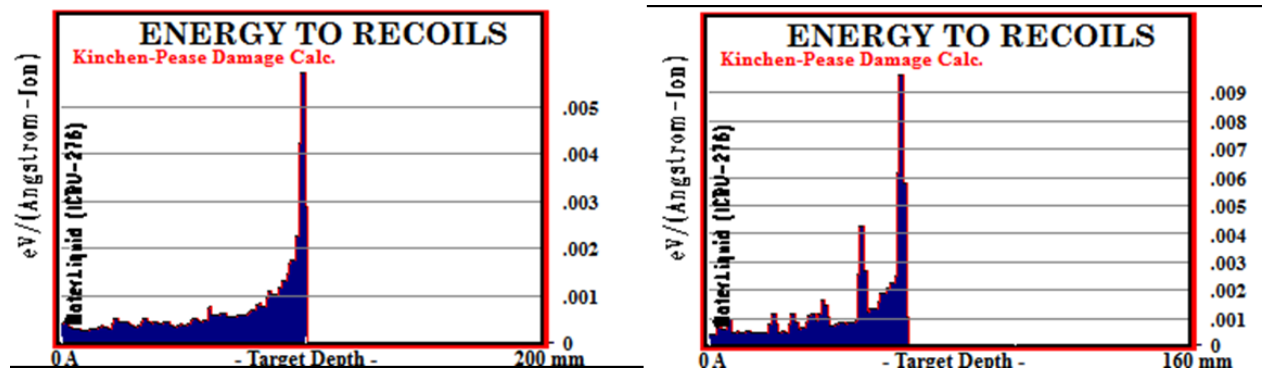

**Figure 2.** Ionization and recoil energies of 2.0 GeV C (left) and B (right) beams

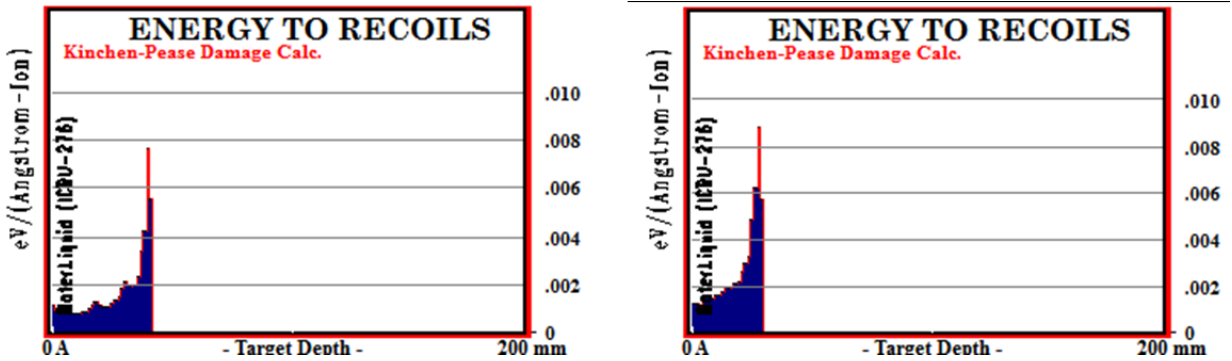

**Figure 3.** Ionization and recoil energies of 2.0 GeV N (left) and O (right) beams

The recoil peaks and ranges in the water phantom for the selected energy range of heavy ions are given in Table 1 and Figure 4-6. It can be seen that the range of heavy ions increases as energy increases. Maximum recoil energy is produced at the end of the range. Boron ion with the lowest mass number has the highest range. As the energy increased, the ranges for B, C, N and O ions increased 0,92 cm, 0,6 cm, 0,36 cm and 0,24 cm, respectively. Near end of range, average recoil energies produced are as follows: B ion: $3.50 \times 10^3$ eV / A-ion, C ion: $5.80 \times 10^3$ eV / A-ion, N ion: $5.35 \times 10^3$ eV / A-ion and O ion: $7.76 \times 10^3$ eV / A-ion.

**Table 1.** Recoil peak values (eV/A-ion×$10^3$), ranges (cm) and percentage differences compared with carbon produced by the beams at the target

| Energy GeV | $^{11}$B Recoil Peak | Range | $^{12}$C Recoil Peak | Range | $^{14}$N Recoil Peak | Range | $^{16}$O Recoil Peak | Range |
|---|---|---|---|---|---|---|---|---|
| 2   | 3.20 | 10.0 | 5.92 | 6.4 | 6.62 | 4.2 | 6.48 | 3.0 |
| 2.1 | 4.01 | 10.8 | 5.65 | 7.0 | 5.88 | 4.6 | 8.41 | 3.2 |
| 2.2 | 2.62 | 11.6 | 5.12 | 7.5 | 5.25 | 5.0 | 9.78 | 3.4 |
| 2.3 | 3.92 | 12.6 | 6.05 | 8.2 | 5.21 | 5.4 | 6.39 | 3.8 |
| 2.4 | 3.87 | 13.6 | 6.36 | 8.8 | 3.58 | 5.6 | 8.67 | 4.0 |
| 2.5 | 3.39 | 14.6 | 5.72 | 9.4 | 5.59 | 6.0 | 6.86 | 4.2 |

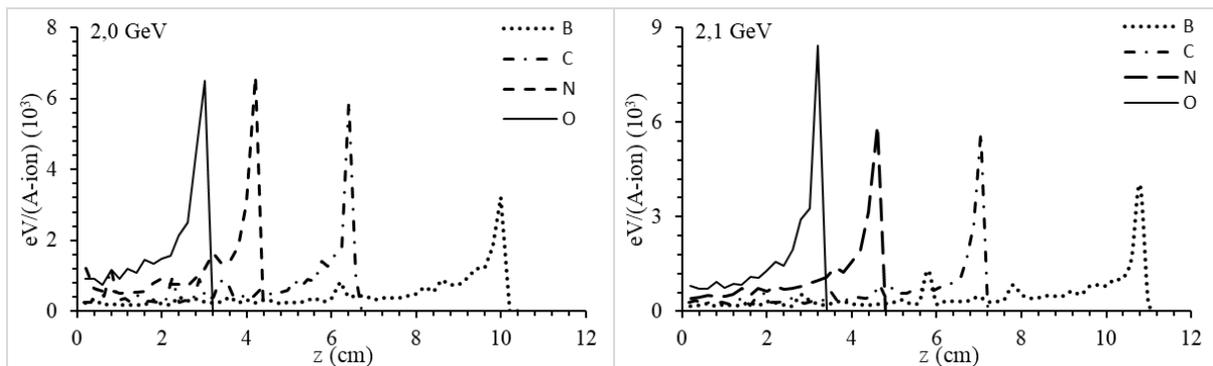

**Figure 4.** Change in the energy absorbed by recoil against depth in the water phantom from 2.0 and 2.1 GeV B, C, N and O beams

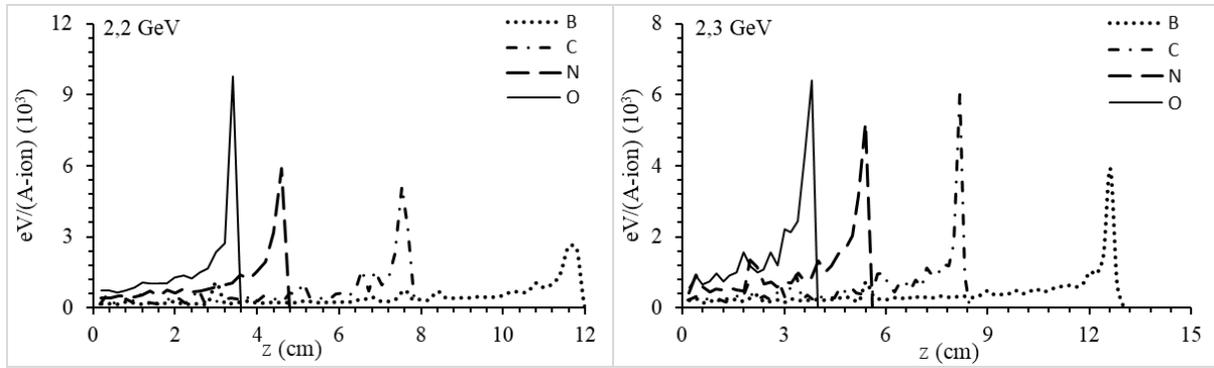

**Figure 5.** Change in the energy absorbed by recoil against depth in the water phantom from 2.2 and 2.3 GeV B, C, N and O beams

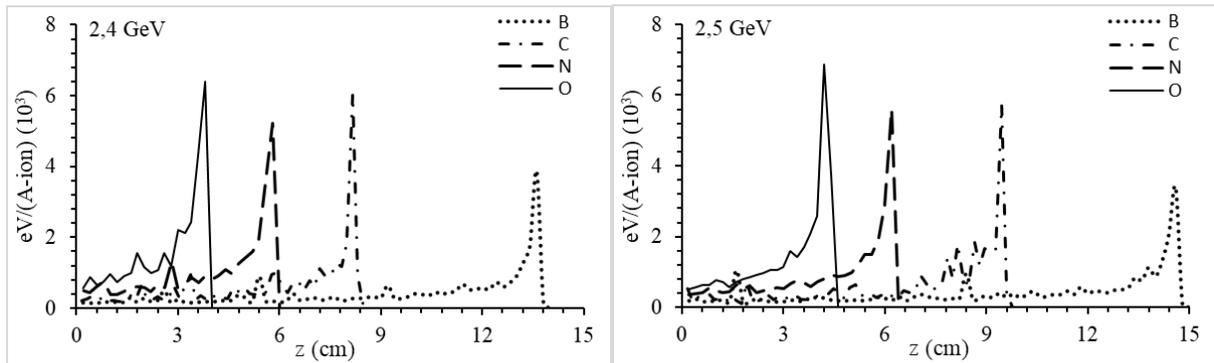

**Figure 6.** Change in the energy absorbed by recoil against depth in the water phantom from 2.4 and 2.5 GeV B, C, N and O beams

The gaps formed by the selected heavy ions in the crystal structure of the water are compared with those of carbon are shown in Table 2. It was observed that as the atomic number of heavy ions increases at the same energy level, the gaps they form in the crystal structure of water increase. We also observed that as the energy of the same heavy ion increases, the gaps in the crystal structure of the water increase. Since, in a water molecule, H has a displacement energy of 10 eV, a binding energy of 3 eV and surface energy of 2 eV and O has a displacement energy of 28 eV, binding energy of 3 eV and surface energy of 2 eV; the majority of the gaps formed are from H atoms. Thus, the H ion (proton) creates ionization and recoils through secondary interactions. In the energy range of 2.0 to 2.5 GeV, B, C, N and O beams created 6392, 6691, 7078 and 7668 gaps on average, respectively.

**Table 2.** Percentage differences of the gaps produced by the ion beams with the gaps produced by carbon

| Energy (GeV) | Heavy ions | | | | % Difference | | |
|---|---|---|---|---|---|---|---|
| | $^{12}C$ | $^{11}B$ | $^{14}N$ | $^{16}O$ | $^{12}C$-$^{11}B$ | $^{12}C$-$^{14}N$ | $^{12}C$-$^{16}O$ |
| 2.0 | 6204 | 5916 | 6573 | 7148 | 4.86 | -5.62 | -13.21 |
| 2.1 | 6402 | 6110 | 6778 | 7360 | 4.77 | -5.55 | -13.01 |
| 2.2 | 6599 | 6301 | 6980 | 7568 | 4.73 | -5.46 | -12.80 |
| 2.3 | 6793 | 6490 | 7180 | 7775 | 4.67 | -5.38 | -12.62 |
| 2.4 | 6983 | 6677 | 7375 | 7979 | 4.58 | -5.31 | -12.47 |
| 2.5 | 7169 | 6858 | 7569 | 8180 | 4.52 | -5.27 | -12.35 |
| **Mean** | **6691** | **6392** | **7076** | **7668** | **4.69** | **-4.43** | **-12.75** |

## 4. Discussion

According to the current literature in biophysics and cancer biology, it is strongly believed that heavy ion therapy should be used to guide the evolution of the therapy in the right direction [23]. There is irrefutable evidence that heavy ion therapy has significant physical, biological and dosimetric advantages over photons and proton beams. Moreover, current clinical evidence indicates promising results in many types of cancer. For this reason, there is a day-by-day increase in the construction of new therapy facilities in both number and geographical location. This increase can also be seen with the research on use of different heavy ions. Studies have focused on linear energy transfer (LET) and relative biological effect (RBE) [24-26]. In these studies, different ion combinations of the C atom have been the focus of attention [27]. The effect of heavy ions were investigated over the neutrons created by the secondary interactions, rather than LET and RBE in [28]. Research should cover the primary and secondary interactions that cause increase in dose [29] and DNA damage [30]. However, research on recoil in this area is generally limited and we have evaluated that it is not at a sufficient level in the field of medical physics [31, 32]. In this study, all interactions of heavy ions were analyzed in order to reveal the recoil interactions in the water phantom.

## 5. Conclusion

As there is an increase in research in heavy ion therapy, an analysis on various effects of using different heavy ions was found as a gap in the literature. This study aimed to fill this gap by providing detailed recoil interactions occurring in the target.

In this study, we compared C beams with B, N and O beams in 2.0-2.5 GeV range. Our findings have shown that C beams have 35.3% more recoil range than B beams, while it has 14.5% and 118.7% less recoil range than N and O beams, respectively. Recoil peak amplitude of C beams is 68.1% more than B beams, while it is 13.1% less than N and 22.9% less than O beams. As the energy increases, there is a gradual increase in recoil peak amplitude in C and B ions, while this pattern is irregular for N and O beams.

Considering the gaps in the crystal structure created by the beams, B beams created 4.69% less gaps than C beams. On the other hand, N and O beams have created 4.43% and 12.75% more gaps than that of C beams. We also observed a general increase in the gaps as the atomic weight and energy increase.

The authors believe that these results will guide the future research in medical physics considering various types of phantoms and biomaterials.